\documentclass[12pt]{article}
\usepackage{graphics}
{\par\begingroup\addtolength{\leftskip}{#1}}%
{\par\endgroup} 
\def\la{\mathrel{\hbox{\rlap{\hbox{\lower4pt\hbox{$\sim$}}}\hbox{$<$}}}}
\def\ga{\mathrel{\hbox{\rlap{\hbox{\lower4pt\hbox{$\sim$}}}\hbox{$>$}}}}

\begin{document}

\begin{center}

{\Large{\bf Water Observed in Red Giant and Supergiant Stars -  
Manifestation of a Novel Picture of the Stellar Atmosphere or else 
Evidence against the 
Classical Model Stellar Photosphere\footnote{Proceedings of the Symposium 
{``\it Exploiting the
ISO Data Archive - Infrared Astronomy in the Internet Age''},
24 - 27 June 2002, Sig\"uenza, Spain (C. Gry et al. eds., ESA SP-511)} 
}}
\vspace{10mm}

\large{Takashi Tsuji} 
\vspace{2mm}
  
{\large{\it Institute of Astronomy, School of Science, The University of 
Tokyo 
Mitaka, Tokyo, 181-0015 Japan
}}

\end{center}

\vspace{16mm}

\noindent
{\bf Abstract:}
We detected the H$_{2}$O 6.3 $\mu$m bands in more than 30 normal red giants  
stars from K5III to M8III as well as in some early M supergiants on the SWS  
spectra retrieved from the ISO Data Archive.
This result, however, shows  serious inconsistency with the present
model photospheres which predict 
H$_{2}$O only in the latest M (super)giant stars.
Also H$_{2}$O was once discovered in the early M (super)giant stars
nearly 40 years ago with the balloon-borne telescope named Stratoscope II.
This discovery was so unexpected at that time that it was not understood
correctly and overlooked for a long time. 
Now, we reflect on our ignorance of this important discovery during the 40
years and should consider more seriously the meaning of the rediscovery 
of water in so many red (super)giant stars with ISO.

\vspace{3mm}


\vspace{20mm}
\begin{flushleft}
{\large{\bf 1. INTRODUCTION}}
\end{flushleft}

At this monumental epoch to celebrate the opening of the ISO active archive 
phase, it may be instructive to recall a short history of stellar 
spectroscopy in space. At the infancy of the infrared astronomy in the 
1960's, an ambitious attempt to observe stellar spectra with a
balloon-born telescope was undertaken and this mission named Stratoscope II,
launched on March 1963, successfully observed infrared spectra (0.8 -- 
3.1\,$\mu$m) of several 
red giant and supergiant stars (and invaluable spectra of Jupiter). 
The results showed beautiful spectra of water in Mira variables 
$o$ Cet and R Leo (Woolf et al. 1964). This result was well in accord
with the theoretical prediction (Russell 1934) and thus was well 
appreciated at that time. 
 
However, the Stratoscope observers reported a more surprising
result that water was detected in the earlier M giants $\mu$ Gem (M3III)
and $\rho$ Per (M4II) as well as in the early M supergiants $\alpha$ Ori
(M2Iab) (Woolf et al. 1964) and $\mu$ Cep (M2Ia) (Danielson et al.
1965). However, this discovery was not in
accord with the understanding of cool stellar atmosphere at that time
and it was finally reinterpreted that the absorption bands at 0.9, 1.1, 1.4, 
\& 1.9\,$\mu$m identified with H$_{2}$O by the Stratoscope observers should
instead be  due to the CN Red System which also has band
heads at about the same positions (Wing \& Spinrad 1970).
This proposition was more easily accepted by the astronomical community
since then, since CN had been observed in a wide range of oxygen-rich 
stars from the Sun to red supergiants, not to speak of carbon stars.

Meanwhile, the discovery of water, at least in the coolest Miras, 
confirmed the importance of water as a source of opacity in cool stars,
and actual computation of the non-grey model photospheres 
revealed that this is true
in red giant stars with $T_{\rm eff}$ cooler than about 3200\,K (Tsuji 1978), 
which roughly corresponds to
M6III (Ridgway et al. 1980). This result was well consistent with the
detection of the strong H$_{2}$O bands in the  Miras but was  
contradicting with the identification of H$_{2}$O in the M2 - M4 
(super)giants by the Stratoscope II observers. Thus, this
result lent further support to the Wing-Spinrad proposition. 
A blind spot  in this apparently reasonable conclusion,
however, was that we were not aware at that time that the model photosphere,
often referred to inadvertently as model atmosphere, is simply a 
model of the {\it photosphere} $(0 \la \tau < \infty)$ and not of the 
{\it atmosphere} $(-\infty  <  \tau < \infty)$, which may still involve 
unknown problems.

Since then, important infrared missions such as IRAS and COBE
have been successfully undertaken, but few observations of the near infrared 
stellar spectra were done in space, except for continued efforts with 
KAO which provided fine low resolution stellar spectra (e.g. Strecker et al. 
1979).  Finally, ISO launched on November 1995 (Kessler et al. 1996) 
provided the means by which to observe astronomical spectra at higher
resolutions for a wider spectral coverage 
at last. One of the unexpected results in the initial ISO observations 
with SWS (de Graauw et al. 1996) was a
detection of water in the M2 giant $\beta$ Peg (Tsuji et al. 1997) and
in several early M supergiants in $h + \chi $ Per clusters (Tsuji et al. 1998).
It took sometime before we recognized that the Stratoscope II observers 
 correctly identified water in the
M2 -- M4 (super)giant stars and that the Wing-Spinrad proposition was not
correct (Tsuji 2000a). 
Further, IRTS launched by ISAS on March 1995  
detected H$_{2}$O bands at 1.9\,$\mu$m in several
M (super)giants earlier than M6 (Matsuura et al. 1999). Also, ground-based
mid-infrared spectroscopy revealed H$_{2}$O pure-rotation lines 
in $\alpha$ Ori and $\alpha$ Sco (Jennings \& Sada 1998).

All these results that H$_{2}$O exists in red (super)giants earlier 
than about M6, however, could not be understood with the present model 
photospheres. We now encounter a serious problem:
are we confronting with a fall of the classical model stellar
photosphere or else with a rise of a novel picture of the stellar atmosphere 
(including the photosphere as a part)?
A rather intriguing story of water in red (super)giant stars
still remains open, and we hope to utilize the extensive ISO Data 
Archive to extend and finalize this fascinating story.

\vspace{10mm}
\begin{flushleft}
{\large{\bf 2. A RED GIANT SAMPLE}}
\end{flushleft}
Our initial  detection of water was done with the H$_{2}$O 2.7\,$\mu$m bands 
which, however, are contaminated by OH, CO, and other molecular bands. 
Then, we analyzed the H$_{2}$O 6.3\,$\mu$m bands which are little disturbed
by other molecular bands. The expected spectra of the H$_{2}$O $\nu_{2}$
bands  computed with the use of HITEMP (Rothman 1997) based on a  single 
absorption slab model  are shown in Fig.1.

We first examined a dozen of high resolution spectra of red giants
in the ISO Data Archive. The 
H$_{2}$O 6.3\,$\mu$m bands are detected in the K5 giant Aldebaran
(but not in other two K giants) as well as in all the M giants between M0III 
and M3.5III (Tsuji 2001), and also in the later M giants (M6-7III),
as shown in Fig.\,2.
We  found $ T_{\rm ex} \approx 1500$K (and log\,$N_{\rm col}$ noted on
Fig.\,2) by referring to the spectra such as shown in Fig.1.
It is remarkable that SWS detected such faint water bands
in the K giant star $\alpha$ Tau. This  result is quite unexpected but 
confirmed recently in another K giant Arcturus ($\alpha$ Boo) with the
high resolution ground-based spectroscopy in the 12\,$\mu$m region 
(Ryde et al. 2002).

Next, we extend our survey to a larger sample of the low resolution
SWS spectra. We found dozens of spectra
in this category from the ISO Data Archive. 
At the lower resolution, some details of the band structure seen at the
higher resolution are smeared out (Fig.1). Nevertheless 
we can detect the dip at 6.63\,$\mu$m due to the H$_{2}$O $\nu_{2}$
bands in 25 M giants from M0III to  M8III. Some examples
are shown in Fig.3 and we estimated  $N_{\rm col}$ values for these 
M giants, again with the single slab model of Fig.1.

The resulting values of $N_{\rm col}$ from the high and low resolution
samples are plotted against spectral types in Fig.4. 
For comparison, the predicted values of $N_{\rm col}$ from the 
spherically extended non-grey model photospheres are shown by the dashed line.
It is clear that the observed $N_{\rm col}$ values cannot be explained 
at all by the predicted ones. Thus, H$_{2}$O detected in M giants 
should be non-photospheric in origin, but where does it come from?

\vspace{10mm}

\begin{flushleft}
{\large {\bf 3. A RED SUPERGIANT SAMPLE}}
\end{flushleft}

In red supergiant stars, 
the H$_{2}$O 6.3 $\mu$m bands are found as absorption  in 
$\alpha$ Ori (M2Iab) and $\alpha$ Sco (M2Ib), as in
K - M giants. Further, water appears in absorption at $\lambda < 5\,\mu$m
but in emission at $\lambda \ga  5\,\mu$m throughout in $\mu$ Cep (M2Ia)
(Tsuji 2000b). This detection of water in distinct emission should be an
important clue to the origin of H$_{2}$O, since such emission should
most probably originate in the outer atmosphere and not, for example, in 
the ``starspots''.

To account for the emission, we upgrade our single slab model
to a spherically extended molecular sphere (MOLsphere for simplicity). 
Since we already know
that   $T_{\rm ex} \approx 1500$\,K and $N_{\rm col} \approx 3 \times 
10^{20}$ cm$^{-2}$ 
from the Stratoscope II data (Tsuji 2000a),
an additional free parameter is the inner radius of the MOLsphere 
$ r_{i}$. For simplicity, we consider only
H$_{2}$O whose absorption cross-section is as large as $ 10^{-18}$ cm$^{2}$
and thus the  H$_{2}$O gas is optically thick. Then we solve the transfer 
equation with the photospheric radiation, resulting in $ F_{\rm P}$,
as a boundary 
condition, which shows absorption bands due to CO, CN, OH, SiO etc (Fig.5).
We found that the resulting emergent flux from the MOLsphere $ F_{\rm P+M}$ 
with $ r_{\rm i} \approx 2 R_{*}$ ($R_{*}$ is the stellar radius)
accounts for the prominent emission lines due to H$_{2}$O at $\lambda \ga   
5\,\mu$m as well as the absorption bands at  $\lambda < 5\,\mu$m
(Tsuji 2000b).

Further, we found that the huge infrared excess can be explained
by an optically thin dust envelope with $\tau^{\rm d}_{10\mu m} \approx 0.1$ 
and $r^{\rm d}_{i} \approx 13.5  R_{*}$.  Then the dust
emission is simply added to $ F_{\rm P+M}$.  
The entire spectra of $\mu$ Cep observed with the ISO SWS (corrected for
$A_{\rm V} = 1.5$\,mag.) can reasonably be 
explained by our final spectrum resulting from Photospher + MOLsphere + Dust
Envelope, $ F_{\rm P+M+D}$. Some details of the water emission 
in the 6\,$\mu$m and 40\,$\mu$m regions can be well reproduced by our model 
as shown 
in the inserted boxes of the left and right, respectively, in Fig.5. 
So far, the huge IR excess of $\mu$ Cep was  thought to be due to 
dust alone, but now it is clear that it includes water emission originating 
in the MOLsphere.
   
\vspace{10mm}

\begin{flushleft}
{\large {\bf 4. DISCUSSION} }
\end{flushleft}
Although we have 
assumed MOLsphere for $\mu$ Cep, this is not a theoretical
model of  the usual meaning but simply a kind of working hypothesis or an
empirical model at best.  We know nothing about the origin of the MOLsphere
and it is a major challenge how to resolve this issue. 
Such a difficulty, however, may be shared with the origin of
the chromosphere as well as of the mass-loss outflow 
(at least in non-pulsating stars).  With this reservation,
our empirical model for $\mu$ Cep is reasonably
successful, and a problem is if such a model can be extended to other
cases.  Although we see no emission in the 6\,$\mu$m region in our large
sample of red giant stars, it is interesting to notice that
the H$_{2}$O column densities  tend to lever-off at about M5 and
then to be smaller than the predicted photospheric values (Fig.4).
Moreover, water bands almost disappear in M7 giant EP Aqr.
These results suggest that the photospheric H$_{2}$O bands may be
filled in by the emission due to H$_{2}$O itself in the late M giants.
Also, in the late M giants, H$_{2}$O appears as emission in the 40\,$\mu$m 
region (Tsuji et al. 1999), CO$_{2}$ shows prominent emission in the
15\,$\mu$m region (Jasttanont et al. 1998), and  SO$_{2}$ shows emission 
as well as absorption in the 7\,$\mu$m region (Yamamura et al. 1999).  
Based on these observations, it should be reasonable to  assume the 
presence of MOLsphere in late M giants.

However, a possibility that there are some serious
flaws in the present model photospheres cannot be excluded, especially
for K (see Ryde et al. 2002) and early M giants for which there is no
direct evidence for MOLsphere.  This case offers
a more serious problem, since our present stellar spectroscopy
(e.g. abundance analysis) loses its basis.  
Also, if this is the case, we must give up to have a unified 
picture for cool luminous stars including K and early M giants.
We are yet tempted to have a unified picture, ISO's innovation of which 
may be illustrated as in Fig.6.   
The presence of the high excited molecular gas has been known for
late (super)giants from the H$_{2}$O  and SiO masers for a long time,
but ISO  revealed that the presence of such a warm molecular gas,
including not only H$_{2}$O  and SiO but also CO$_{2}$ and SO$_{2}$, 
may be a general phenomenon in red giant and supergiant stars.

\vspace{10mm}

\begin{flushleft}
{\large {\bf 5.  CONCLUDING REMARKS} }
\end{flushleft}

In concluding, we summarize our present viewpoint:
\begin{enumerate}

\item  Water  was discovered in several early M (super)giants 
nearly 40 years ago with    
Stratoscope II, but this  discovery has been 
mis-interpreted and overlooked until recently. We regret
that this important discovery  did not  provide 
major impact on the  theory of cool stellar atmosphere during
these 40 years. We certainly hope that such an unfortunate 
history should not be repeated with the ISO data.
 
\item  After 40 years, we confirmed H$_{2}$O absorption bands in more than 30 
red giant stars from K5III to M8III with the use of the ISO Data Archive. Thus
ISO finally established that the presence of water  is 
a general phenomenon in red  giant stars including K and early M types. 

\item  With ISO SWS, we detected water
in emission in the M2 supergiant $\mu$ Cep. This fact suggests
that water should be in the outer atmosphere 
rather than in the photosphere. Also, emission not only of H$_{2}$O 
but also of CO$_{2}$ and SO$_{2}$ are detected in
late M giants with ISO and, by implication, water in all the red giant
stars may also be originating in the outer atmosphere.

\item We conclude that the presence of a rather warm molecular sphere --
MOLsphere -- may be a general feature in red giant and
supergiant stars. Thus ISO revealed a novel picture of the atmosphere
consisting not only of the photospher, chromosphere, and wind so far
known but also of a new component -- MOLsphere. Certainly, the ISO Data 
Archive should be an invaluable tool by which to explore the fundamental 
problem on the atmospheric structure of red giant and supergiant stars.

\end{enumerate}

\vspace{15mm}
\noindent
{\bf Acknowledgements:}
I thank I. Yamamura and T. Tanab\'e for helpful discussion on the 
ISO data analysis. 
This work, supported by JSPS 
grant No. 11640227, was carried out with the facilities at NAO ADAC.

\newpage
\begin{figure}[t]
  \begin{center}
\resizebox{11.6cm}{!}{
\rotatebox{0}{\includegraphics{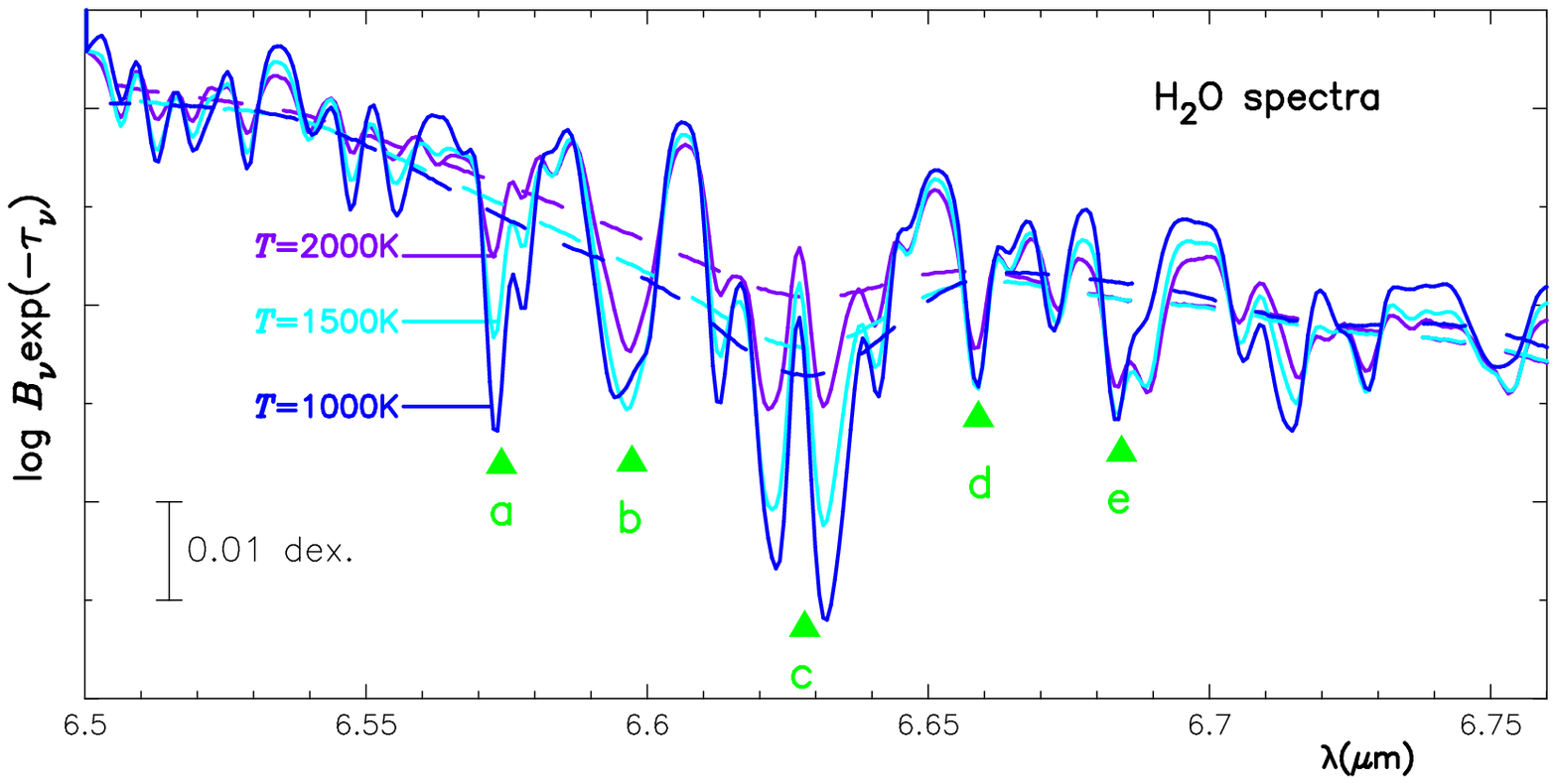}}}
  \end{center}
\caption{\it
Spectra of water for $N_{\rm col} = 10^{18}$ cm$^{-2}$ and 
 $T_{\rm ex} = 1000, 1500,$ and $2000$\,K are shown for high 
($R \approx 1600$; solid lines) and low ($R \approx 200$; dashed lines) 
resolutions of SWS.
}
\end{figure}

\begin{figure}[b]
  \begin{center}
\resizebox{11.6cm}{!}{
\rotatebox{0}{\includegraphics{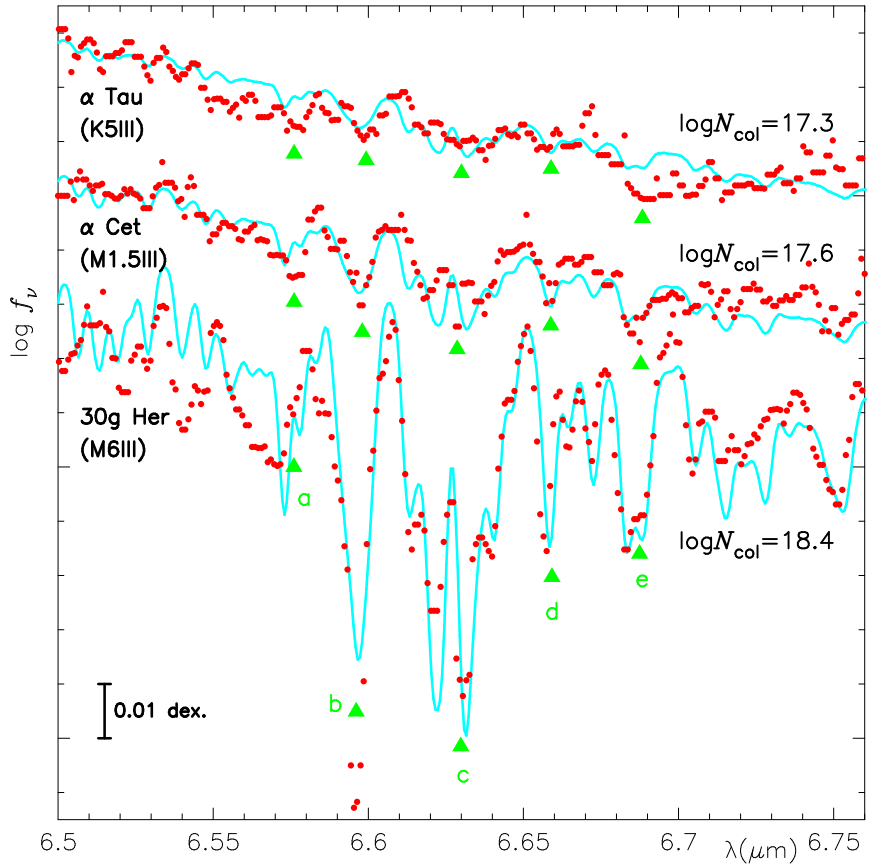}}}
  \end{center}
\caption{\it
 Observed high resolution  SWS spectra (filled circles) 
are compared with the water spectra of Fig.1 (solid lines).
}
\end{figure}

\begin{figure}[t]
  \begin{center}
    \resizebox{11.6cm}{!}{
    \rotatebox{0}{\includegraphics{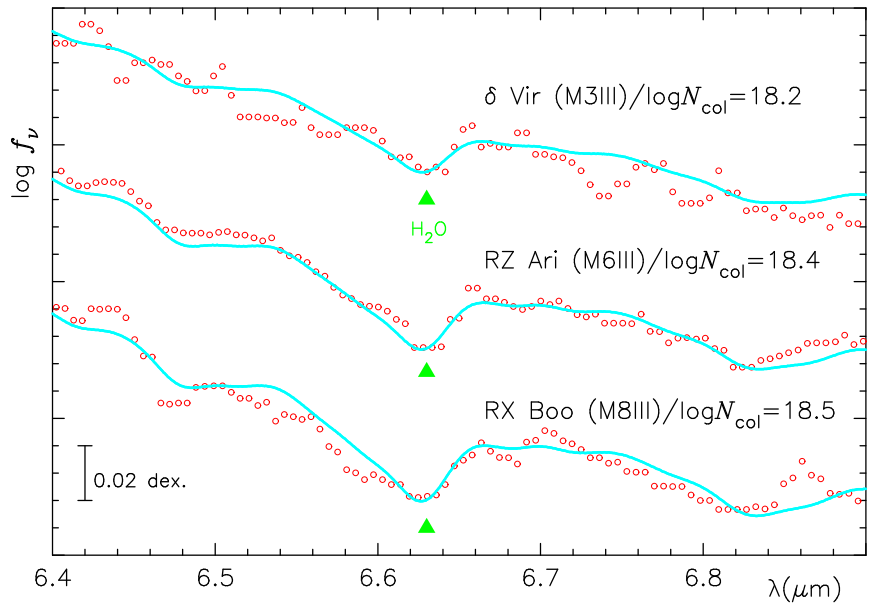}}}
  \end{center}
\caption{\it
 Observed low resolution  SWS spectra (open circles) 
are compared with the water spectra of Fig.1 (solid lines).
}
\end{figure}

\begin{figure}[hb]
  \begin{center}
    \resizebox{11.6cm}{!}{
    \rotatebox{0}{\includegraphics{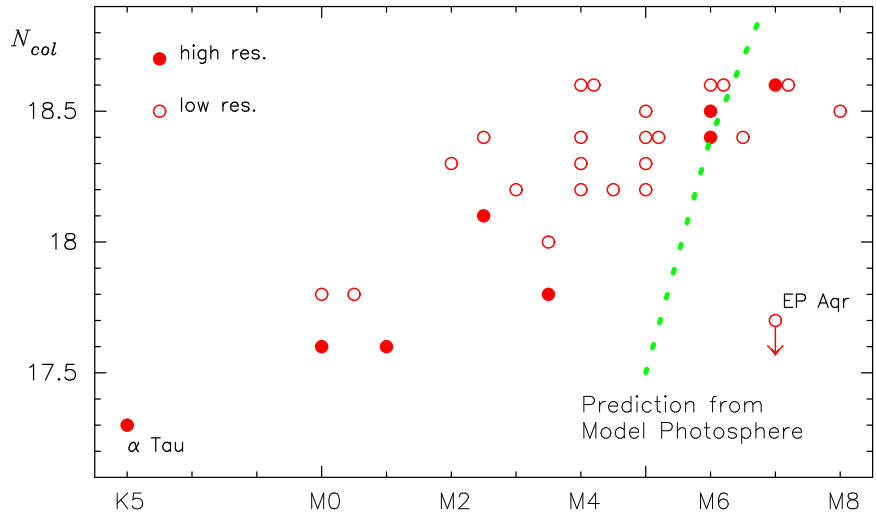}}}
  \end{center}
\caption{\it
The observed $N_{\rm col}$ values from  the high and low
resolution SWS spectra (filled and open circles, respectively) plotted
against spectral types are compared with the predicted $N_{\rm col}$ values
based on the model photospheres (dashed line).
}
\end{figure}

\begin{figure}[ht]
  \begin{center}
    \resizebox{13.5cm}{!}{
    \rotatebox{0}{\includegraphics{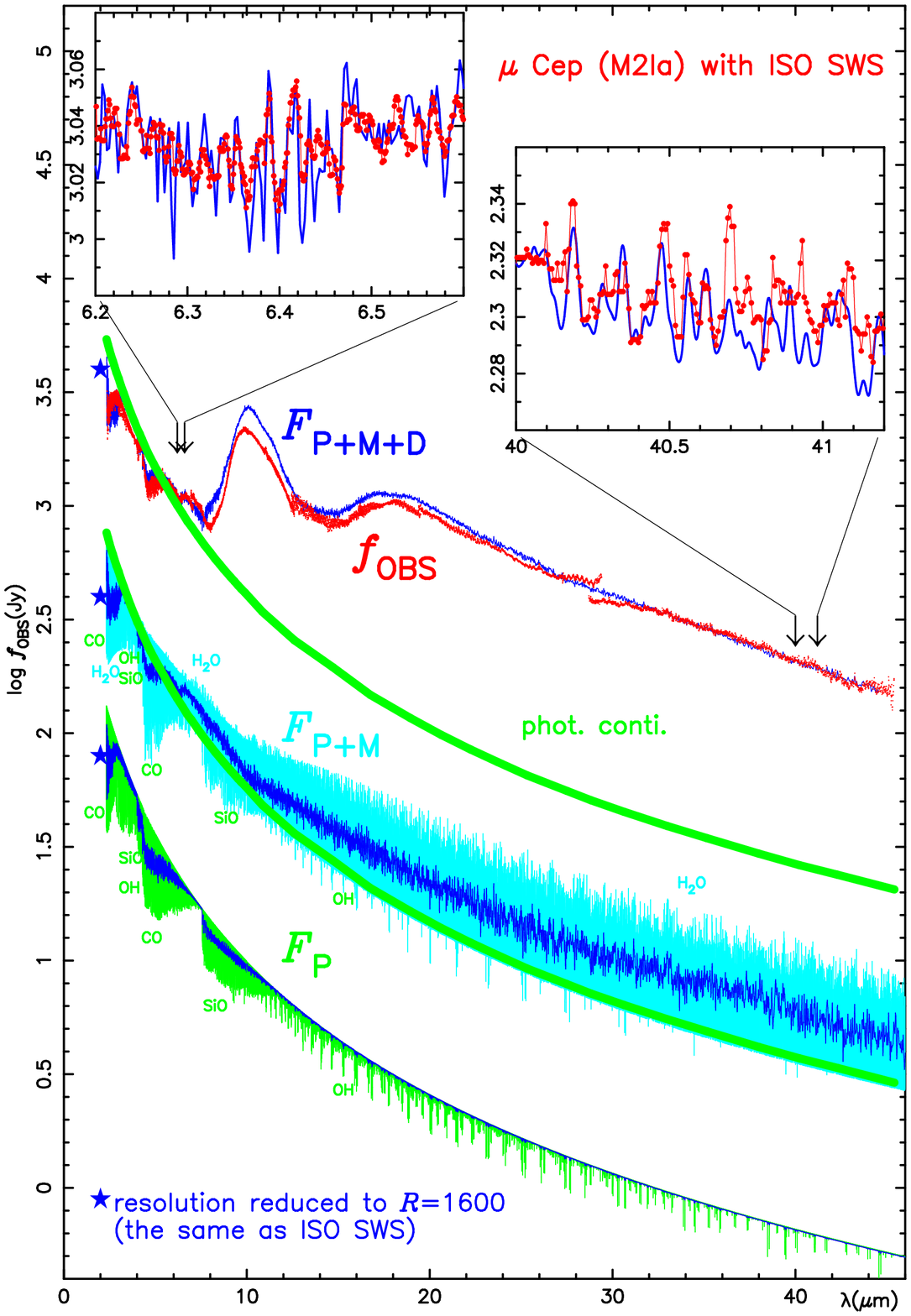}}}
  \end{center}
\caption{\it
Photospheric spectrum $F_{\rm P}$ (bottom) based on the spherically
symmetric non-grey model photosphere ($T_{\rm eff} =3600$\,K, 
$M_{*} = 15\,M_{\odot}$, $R_{*} = 650\,R_{\odot}$) is used as the 
boundary condition for solving radiative transfer in the MOLsphere.
 The resulting emergent spectrum from the MOLsphere $F_{\rm P+M}$ (middle)
is further diluted by the emission due to the optically thin dust envelope 
and  $F_{\rm P+M+D}$ (top)  
is the final spectrum to be compared with the
ISO spectrum $f_{obs}$.
The computations of $F_{\rm P}$ and $F_{\rm P+M}$ are done with a 
resolution of $R \approx 10^{5}$ and the results are convolved 
with the slit function of SWS ($R = 1600$). 
}
\end{figure}

\begin{figure}[ht]
  \begin{center}
    \resizebox{11.6cm}{!}{
    \rotatebox{0}{\includegraphics{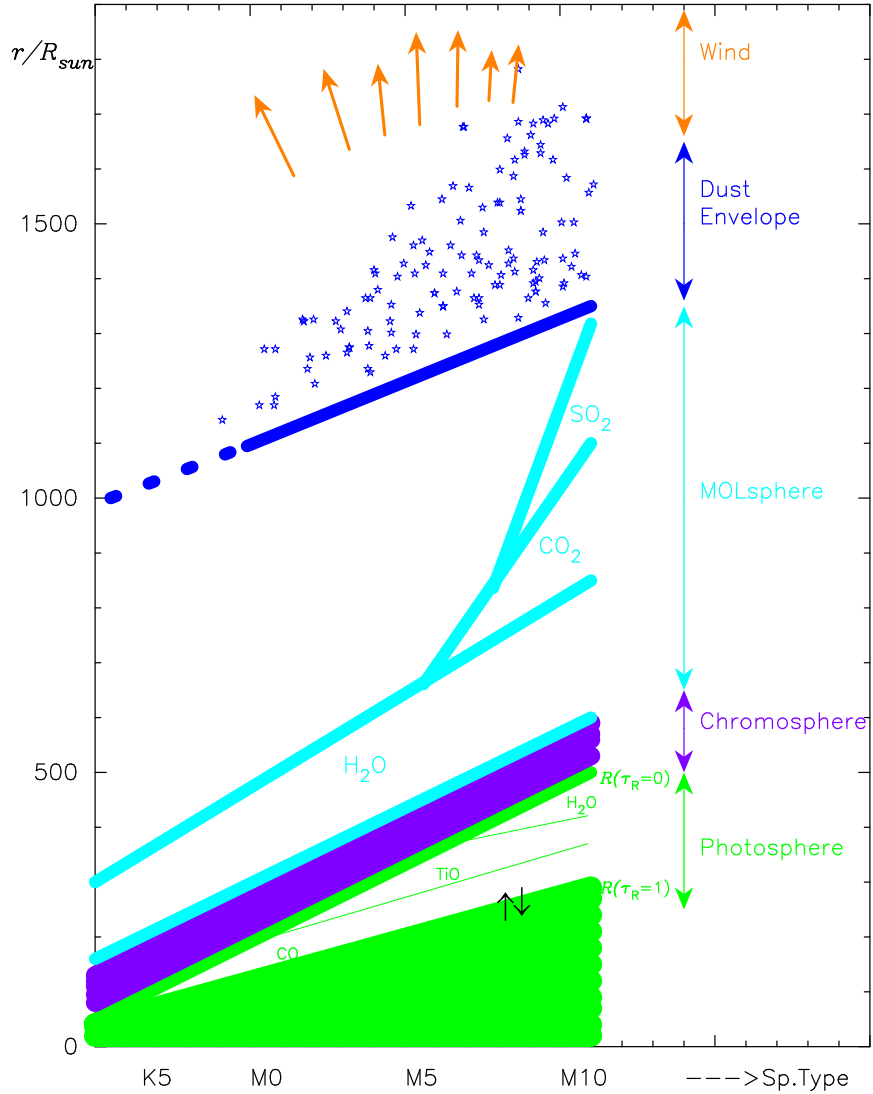}}}
  \end{center}
\caption{\it
A working hypothesis on the evolution of the atmospheric  structure of 
red giants with Sp.\,Type.
Stellar radius $R_{*}$ is defined by 
$\tau_{\rm Ross} \approx 1$ but photosphere 
extends to where $\tau_{\rm Ross} \approx 0$.  
Presently, self-consistent modelings are possible for the photosphere 
and interior ( $ 0 < \tau  < \infty $),
but not at all for other components in the outer atmosphere 
($ -\infty < \tau < 0$). 
}
\end{figure}

\end{document}